\documentclass[preprint]{aastex}

\tighten

\accepted{12 November 1999}

\slugcomment{To appear in 2000 Apr 10 issue of ApJ.}

\newcommand {\sqig} {$\sim$}
\newcommand {\heao} {{\it HEAO-1}}
\newcommand {\asca} {{\it ASCA}}

\newcommand {\rosat} {{\it ROSAT}}

\newcommand {\sax} {{\it BeppoSAX}}
\newcommand {\src} {X1832$-$330}
\newcommand {\clu} {NGC~6652}

\shorttitle{LMXB in NGC 6652}
\shortauthors{Mukai \& Smale}

\begin{document}

\title{The Low-Mass X-ray Binary X1832$-$330 \\
in the Globular Cluster NGC 6652: \\
A Serendipitous ASCA Observation \\ }

\author{Koji Mukai\altaffilmark{1} and Alan P. Smale\altaffilmark{1}}
\affil{Code 662, NASA/Goddard Space Flight Center, Greenbelt, MD 20771, USA.}

\altaffiltext{1}{Also Universities Space Research Association}

\begin{abstract}

The Low Mass X-ray Binary (LMXB) \src\ in \clu\ is one of 12 bright,
or transient, X-ray sources to have been discovered in Globular Clusters. 
We report on a serendipitous {\sl ASCA\/} observation of this Globular
Cluster LMXB, during which a Type I burst was detected and the persistent,
non-burst emission of the source was at its brightest level recorded to date.
No orbital modulation was detected, which argues against a high inclination
for the \src\ system.  The spectrum of the persistent emission can be fit
with a power law plus a partial covering absorber, although other models
are not ruled out.  Our time-resolved spectral analysis through the burst
shows, for the first time, clear evidence for spectral cooling from
kT=2.4$\pm$0.6 keV to kT=1.0$\pm$0.1 keV during the decay.  The measured
peak flux during the burst is $\sim$10\% of the Eddington luminosity for
a 1.4M$_\odot$ neutron star.  These are characteristic of a Type I
burst, in the context of the relatively low quiescent luminosity of \src.

\end{abstract}


\section{Introduction}

Of the \sqig 150 globular clusters associated with the Milky Way galaxy,
12 have been seen to harbor a bright ($L_x > 10^{36}$ ergs\,s$^{-1}$),
or transient, low-mass X-ray binary (LMXB)
\citep{v95}.  These binaries are presumably formed through
stellar encounters in the dense cores of the clusters; such events play
an important role in the dynamical evolution of the clusters, as the formation
of a single LMXB can impart enough kinetic energy to the surrounding stars
to terminate a core collapse.  At the same time, the globular cluster LMXBs
provide a unique opportunity to study LMXBs at a well-known distance with
a well-known (and usually very poor) metalicity level.

The X-ray source \src\ in the globular cluster \clu\ is a lesser known example
of this class.  Although the error box for the \heao\ source, H1825$-$331,
contained this cluster, it was originally not considered to be a secure
identification because the error box covered a 2.7 deg$^2$ area in Sagittarius
\citep{h85}.  The first secure detection of \src\ as a globular cluster LMXB
was made during the course of the \rosat\ all-sky survey \citep{p91}. 
More recently, it was detected in pointed \rosat\ observations \citep{j96},
and two Type I X-ray bursts from this source, as well as
the persistent emission, have been detected using the Wide Field Camera of
the \sax\ satellite \citep{i98}.  Thus there is now
a strong circumstantial evidence that the \heao\ source H1825$-$331 and
\src are the same source; here we have adopted this identification as a
working assumption.

In the abovementioned papers, the distance to \src\ was assumed to
be \sqig 14.3 kpc.  However, the first published color-magnitude diagram of
this cluster \citep{o94} has led to the
re-evaluation of the distance to \sqig 9.3 kpc, based on the measured V
magnitudes V$_{\rm HB}$=15.85$\pm$0.04 of its horizontal branch (as well as the
interstellar reddening, E$_{\rm (B-V)}$=0.10$\pm$0.02).  Moreover,
\clu\ appears to be significantly younger than the average globular
clusters \citep{c96}.
Thus the LMXB \src\ in \clu\ may provide an important comparison with other
globular cluster sources, due to the relative youth and the relatively
high metal content ([Fe/H]=$-$0.96) of this cluster (though not the highest
among globular clusters with LMXBs).

A search for the optical counterpart has recently been carried out
using new ground-based data along with archival HST data \citep{d98}.
Although the archival HST observations do not
completely cover the X-ray error circle, the most promising candidate
for the optical counterpart is their star 49, which is relatively faint
($M_B$=+5.5) compared with those of other globular cluster LMXBs.

In this paper, we present our analysis of a serendipitous \asca\ observation
of \src; in comparing the previous observations with the \asca\ data,
we have recalculated the previously-published
source luminosities for a new fiducial distance of 9.3 kpc.

\section{Observation}

The region of the sky containing \clu\ was observed with the
Japanese X-ray satellite, \asca\ \citep{t94}
between 1996 Apr 6 20:12UT and Apr 7 19:00 UT (seq no 54016000).  This
observation was part of a program to observe diffuse Galactic emission,
and only serendipitously included \src\ in its field of view.
There are four co-aligned X-ray telescopes on-board \asca, two with
SIS (Solid-state Imaging Spectrometers, using CCDs) detectors and two with
GIS (Gas Imaging Spectrometers) detectors.  However, little useful data were
taken with GIS-2, due to a problem with its on-board processor\footnote{This
problem had been noticed during the ground contact around 1996 Apr 5
20UT but initial attempt to fix this was unsuccessful.  After the software
was reloaded, the CPU returned to normal status at Apr 7 18:12UT.  There is
therefore \sqig 40 min of useful GIS-2 data, which we have chosen not to
analyze.}.

No such problems exist for the SIS data; however, the observation
was done with both SIS detectors in 4-CCD mode, with \src\ near the center
of the field of view.  This pointing direction minimizes the vignetting,
but the photons from \src\ are spread over all 4 chips, complicating the
analysis.  Moreover, 4-CCD observations suffer most severely
from the cumulative effect of radiation damage.  As a consequence,
events below \sqig 0.7 keV had to be discarded on-board to avoid
telemetry saturation due to flickering pixels, and the spectral resolution
and the quantum efficiency are both severely degraded \citep{d95}.
The degradation is believed to be due largely to residual dark distribution
(RDD): a significant fractions of the CCD pixels now show elevated levels
of dark current, the histogram of which is strongly skewed.  When RDD-affected
data are processed (on-board or on the ground) assuming a Gaussian
distribution of dark current, this leads to incorrect pulse heights
or spurious rejection of X-ray events as particle events.
The current version of the response generator has a model of the
degrading spectral resolution, but not one of the degrading quantum efficiency. 
We have used the FTOOL, {\tt correctrdd}, which partially recovers
the detection efficiency; however, this algorithm is not 100\% effective.
Moreover, the calibration of RDD-corrected data is uncertain.  Therefore,
we have primarily relied on the GIS data, cross-calibrated the RDD-corrected
SIS data against the GIS data, and used the SIS data only when GIS data
were unavailable.

Since a bright source is clearly detected (see below), we have opted to use
loose sets of screening criteria.  For the GIS, we use non-SAA, non-Earth-occult
(ELV$>5^\circ$) data at the standard high-voltage setting, and only exclude
regions where the cut-off rigidity is less than 4 GeV/c; note that, for
safety reasons, the high voltage is reduced well before the satellite
enters the SAA.  After screening, we are left with \sqig 42 ksec of good
GIS-3 data.

For the SIS, we use additional criteria that the line of sight must be
$>$20$^\circ$ away from the sunlit Earth, the time after day/night and
SAA transitions must be $>$128 s, and the PIXL monitor counts for
the CCD chips must be within the 3$\sigma$ of their mean value.
Moreover, we have imposed the condition that the data must have been taken
in the Faint mode.  To correct for RDD and DFE (dark frame error, the
variation in the mean dark level of all the pixels due primarily to scattered
light on the CCD), we have applied {\tt faintdfe} to the original Faint mode
data first, followed by {\tt correctrdd}, before converting to Bright2 mode,
to minimize the interference between DFE and RDD corrections).  This resulted
in \sqig 21 ksec of good SIS data.

We have tested the calibration of RDD-corrected SIS data, by performing
simultaneous fits to the GIS-3 and SIS data.  We find that, even after
the RDD correction, the best-fit SIS model contains a spurious excess
N$_H$ of 1.6$\times 10^{21}$ cm$^{-2}$ as well as a normalization below
that of the GIS-3 data by a factor of 1.17.

\section{Results}

\subsection{GIS-3 Data}

\begin{figure}[th]
\begin{center}
\plotone{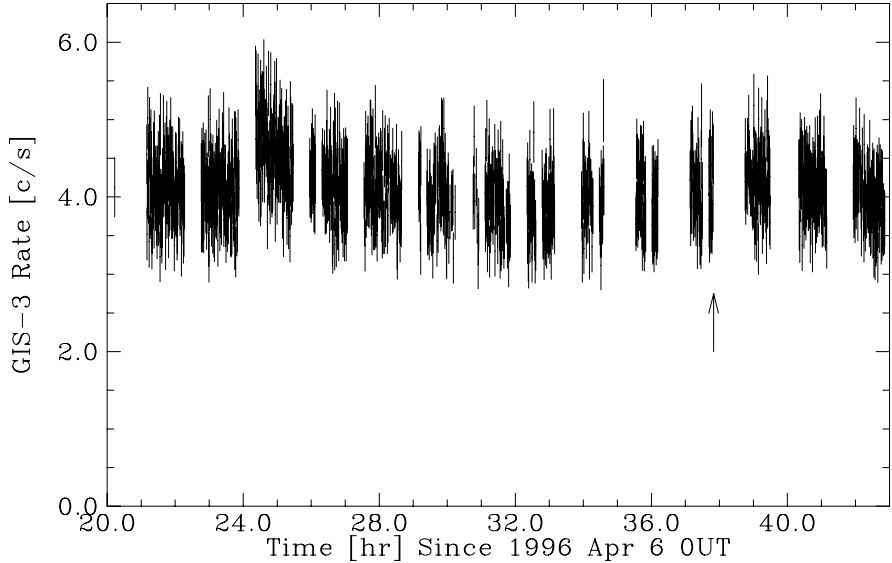}
\caption{The GIS-3 light curve of \src, in 128 s bins.
The arrow indicates the time of the burst, which occurred at the begging
of a GIS-3 data gap.}
\end{center}
\end{figure}

In Fig.1, we have plotted the background subtracted light curve of
\src\ in 128 s bins.  The source appears variable on short timescales
(from about a few bins of this diagram down to 4 sec): a straight line
fit to a 4-s bin light curve, after removing the longer-term trend
(by subtracting a 256-s running average of itself) yields a $\chi^2_\nu$
of 1.114 for 10479 degrees of freedom, meaning that the source is
variable at a formal confidence level of $1- (1 \times 10^{-15}$).
However, some caution is required at this level: although background
is negligible, there may be systematic contribution to this apparent
variability from, e.g., attitude jitter or the imperfect correction
of the time-dependent detector gain.   Moreover, a Fourier transform
did not reveal a periodicity in the range 8 s to \sqig 1 hr (with an
RMS amplitude of \sqig 0.5\% in this range); the highest peaks in the
periodogram in this range have semi-amplitudes of 1.6 \%, while a signal
would have to have an amplitude of $>$2 \% to be detected at $>$99\%
confidence.

Although there are some possible peaks in the periodogram at longer periods,
we consider these to be rather unreliable, since they can be explained
as due to an interplay of the quasi-regular data gaps and the increased
count rate between 0 and 2 UT on Apr 7 (see Fig. 1); this flare-like
event may well be part of the aperiodic variability.   We do not see
spectral changes during this flare-like event (such as might be expected
were it the tail of a Type I burst).  The highest peaks in
the periodogram are at P=46600s ($>$half the duration of the observation)
with a 4.5\% amplitude, and at 17400 s (2.8\%).  Although non-sinusoidal
modulation (e.g., dipping activities) with certain periods (e.g., near the
96 min spacecraft orbit) may have eluded detection, this would seem to
require an unfortunate coincidence.

\begin{figure}[th]
\begin{center}
\plottwo{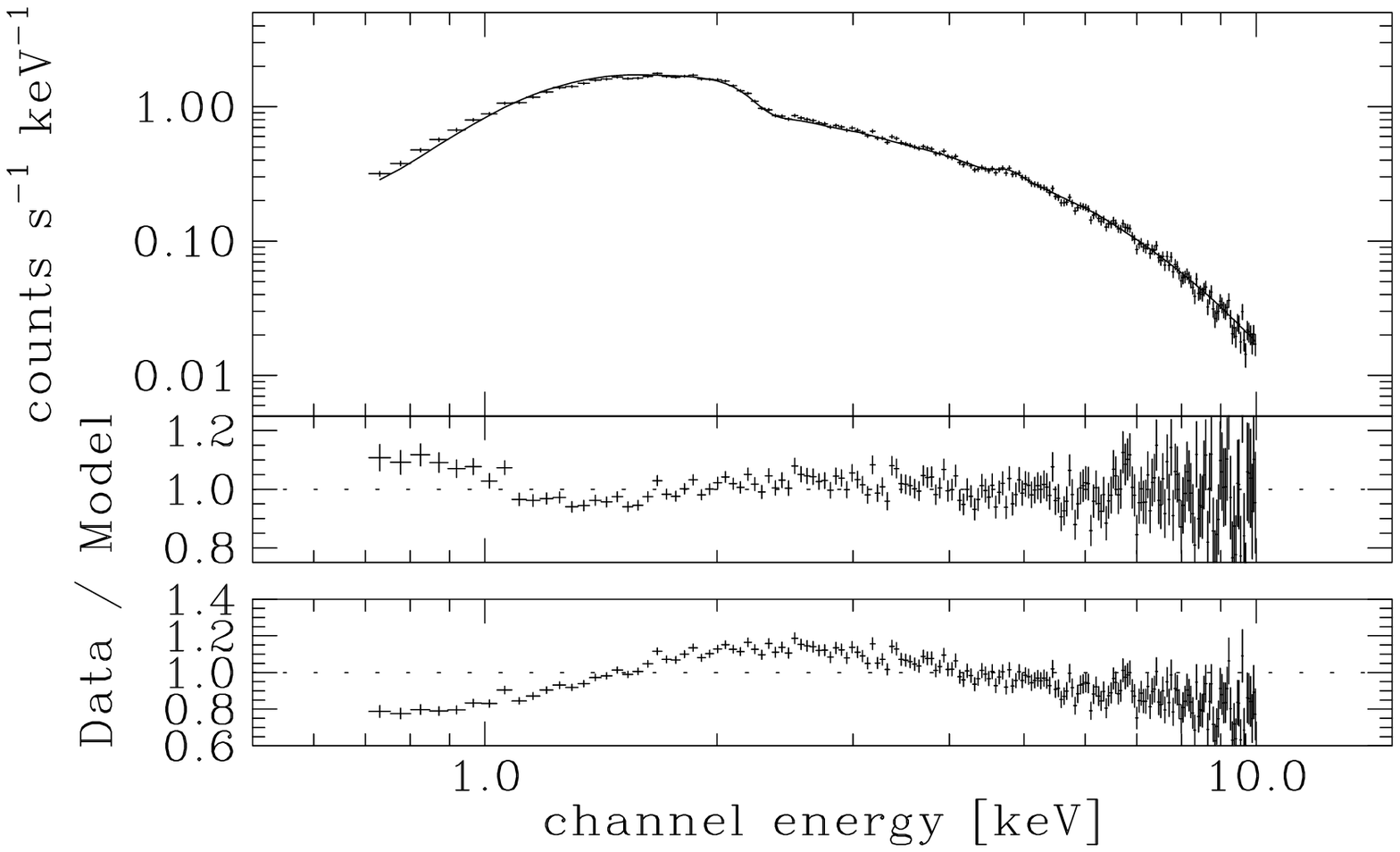}{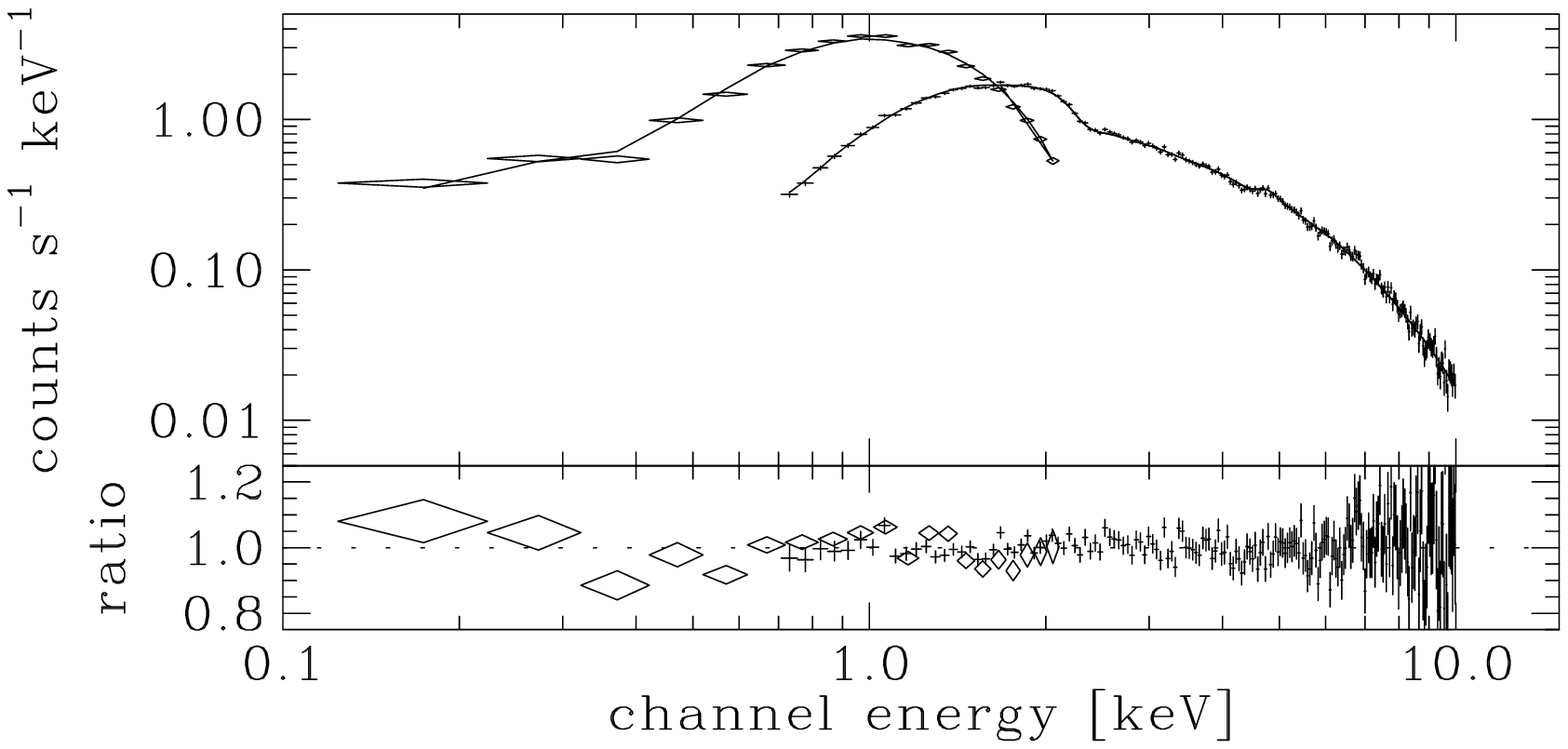}
\caption{(a) The average GIS-3 spectrum of \src,
plotted with the best-fit power-law model (top), with the residuals shown
in the form of the data/model ratio (middle).  In the bottom panel, we
show the residual for the best-fit power-law model, when N$_H$ was fixed
to 6.7 $\times 10^{20}$ cm$^{-2}$, as found for \rosat\ PSPC spectrum.
(b) The \asca\ GIS-3 and \rosat\ PSPC spectra of \src, fitted simultaneously
using a power law model ($\alpha$=1.84$\pm$0.02) modified with
a partial-covering absorber (N$_H$=7.6$^{+0.5}_{-0.6} \times 10^{21}$
cm$^{-2}$, covering fraction 0.60$^{+0.2}_{-0.1}$), also with an interstellar
absorption model of N$_H$ = 8.9$^{+0.3}_{-0.4} \times 10^{20}$ cm$^{-2}$.}
\end{center}
\end{figure}

In Fig. 2(a), we present the average GIS-3 spectrum of \src\ with the
best-fit power-law model.  The fit is poor, with $\chi^2_\nu$ = 1.81;
the parameters are photon index $\alpha$ = 1.75$\pm$0.01 and
N$_H$=3.6$^{+0.2}_{-0.1} \times 10^{21}$ cm$^{-1}$, and a 2--8 keV flux
of 1.54$\times 10^{-10}$ ergs\,cm$^{-2}$s$^{-1}$. 
Note that the fitted N$_H$ is considerably greater than that
estimated from the optical extinction ($\sim 5 \times 10^{20}$ cm$^{-2}$)
or the value derived from the \rosat\ PSPC spectral fit
($6.7\pm 0.2 \times 10^{20}$ cm$^{-2}$; see also the bottom panel
of Fig. 2(a)).  Moreover, the inferred photon index
$\alpha$ is radically different between the \asca\ GIS (1.75) and \rosat\ PSPC
(1.07) observations.  We therefore must conclude that the spectrum of \src\
is highly variable, more complex than a simple power law, or both.

As a likely candidate for the complex spectral shape, we have tried
a power law modified by a partial covering absorber (with a fixed interstellar
absorption of $\sim 5 \times 10^{20}$ cm$^{-2}$) to the GIS-3 data.  This has
markedly improved the fit (to $\chi^2_\nu$=1.15;
$\alpha$=1.86$^{+0.03}_{-0.02}$, with \sqig 67\% covering by a
N$_H$=7.6$\pm 0.9 \times 10^{21}$ cm$^{-2}$ absorber).  Moreover, this model
provides a plausible description of the spectrum in a simultaneous fit to the
\asca\ GIS and \rosat\ PSPC spectra (Fig. 2(b)).  We therefore conclude
that the X-ray spectrum of \src\ is not a simple power law.  However,
given the energy range of the data, and the current level of calibration
uncertainties, we cannot say for sure if this description of the spectral
shape is unique, or preferred.

\begin{table}
\begin{center}

Table 1. Long-term Variability of X1832$-$330. \\

\begin{tabular}{llll}
\hline\hline
Epoch & Instrument & 2--8 keV Flux$^a$ & 2--8 keV Luminosity$^b$ \\
\hline
1977/1978 & {\sl HEAO-1\/} A1$^c$
	& 8.3 $\times 10^{-12}$ & 8.6 $\times 10^{34}$ \\
1992 Apr & {\sl ROSAT\/} PSPC
	& 1.16 $\times 10^{-10}$ & 1.2 $\times 10^{36}$ \\
1996 Apr & {\sl ASCA\/} GIS & 1.54 $\times 10^{-10}$ & 1.6 $\times 10^{36}$ \\
1996 \& 1997 & {\sl BeppoSAX} WFC & 5.4 $\times 10^{-11}$
	& 5.6 $\times 10^{35}$ \\
\hline
\end{tabular}
\end{center}

$^a$Measured or inferred flux in the 2--8 keV band, in
	ergs cm$^{-2}$s$^{-1}$. \\
$^b$Inferred 2--8 keV luminosity in erg s$^{-1}$ for an
	assumed distance of 9.3 kpc. \\
$^c$Source identification remains tentative. \\

\end{table}

In Table 1, we have summarized the long-term history of the X-ray luminosity
of \src.

\subsection{SIS Data}

For the time intervals when both GIS-3 and the SIS instruments were taking
data, the latter adds little.  However, we have discovered a Type I X-ray
burst from \src\ in the section of SIS data for which there is no GIS-3
coverage.  The light curves in three energy bands are shown in Fig. 3(a).
The reason why this burst was not covered by the GIS-3 data is that this
happened just before \asca\ went into the SAA; the high voltage level
of the GIS had already been reduced as a precaution.  This segment of
data ends when the SIS also stopped taking data, as \asca\ approached
the SAA.  We have therefore examined the housekeeping as well as scientific
data carefully to ascertain that this event is not an instrumental artefact.
However, the radiation belt monitor counts indicate that the particle
background was $<$10 ct\,s$^{-1}$, i.e., at quiescent (non-SAA) level
(all data with monitor rates up to 500 ct\,s$^{-1}$ have routinely been
included in GIS data analysis, with no obvious ill effects).  The monitor
rate exceeded 1000 ct\,s$^{-1}$ \sqig 200 s after the end of the SIS data
(in contrast, the radiation belt monitor rates exceed 10,000 in the
heart of the SAA).  Moreover, the image of the burst is identical, to
within statistics, to the quiescent image (i.e., has a distribution
consistent with the XRT point spread function).  Thus, we believe
that the burst originates from the same point-like source as the quiescent
emission, i.e., most likely \src\ in \clu.

\begin{figure}[th]
\begin{center}
\plottwo{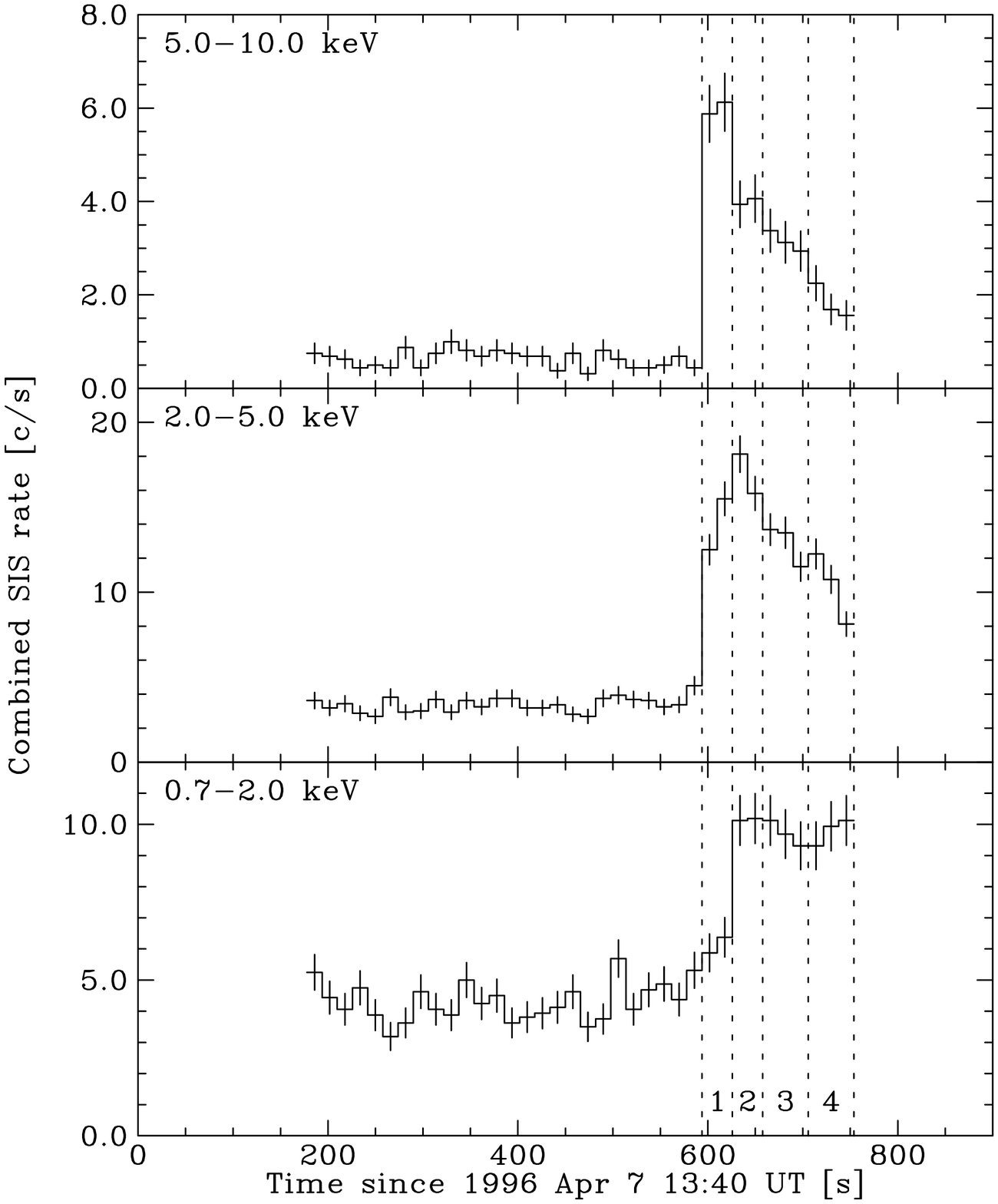}{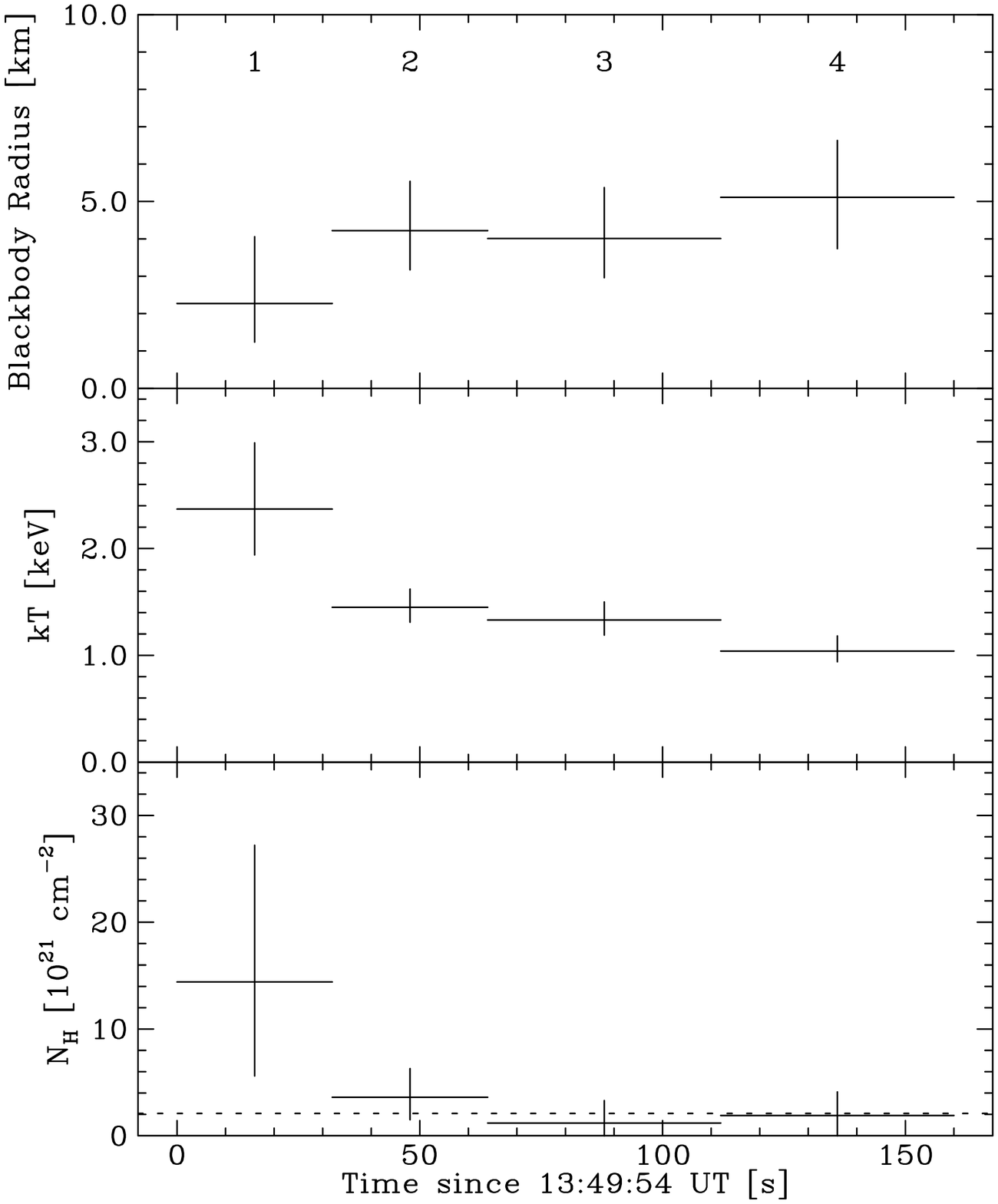}
\caption{(a) The SIS light curve of the X-ray
burst, in 3 energy bins.  Time intervals used for the spectral fitting
are also indicated. (b) The results of the burst spectral fits.}
\end{center}
\end{figure}

The longer duration at lower energies, shown in Fig. 3(a), is what is
expected in a Type I X-ray burst, as the neutron star cools.  To further
investigate the spectral evolution, we have performed spectral fitting of
the 4 time intervals indicated in Fig. 3(a).  We have used the combined
SIS-0/SIS-1 data, and the quiescent SIS spectrum as the background.
We present the results of blackbody fits in Fig. 3(b).  For interval 1,
we find that a significant N$_H$ is required to fit the data adequately;
for the other intervals, the fitted N$_H$ values are consistent with the
interstellar N$_H$ (\sqig 5$\times 10^{20}$ cm$^{-2}$), once the systematic
offset of 1.6$\times 10^{21}$ cm$^{-2}$ (see \S 2) has been taken into account.

As is typical of Type I bursts, the color temperature shows a significant
decline during the decay of the burst.  The inferred radius of the blackbody
emitter (we have used the distance of 9.3 kpc and included the normalization
correction factor of 1.17) also shows behavior typical of Type I bursts,
although it may be on the small side.  The inferred bolometric flux during
interval 1 is 2.03$\times 10^{-9}$ ergs\,cm$^{-2}$s$^{-1}$, thus the bolometric
luminosity is 2.1$\times 10^{37}$ ergs\,s$^{-1}$.  This may underestimate
the true peak flux/luminosity somewhat, due to the limited time resolution
of our data; judging by the light curve, the true peak values are unlikely
to be greater by $>$1.5 compared with the interval 1 averages.  The burst
fluence (integrated over the 160s interval for which we have data) is
estimated to be 1.45$\times 10^{-7}$ ergs\,cm$^{-2}$ (equivalently,
total burst energy of 1.45$\times 10^{39}$ ergs); the fact that we did
not see the return to quiescence may have resulted in underestimating
this by \sqig 10\%.  Thus the average duration $\tau$ of the burst was
\sqig 71 s.

\section{Discussion}

The quiescent X-ray spectrum of \src\ appears to have a complex shape.
A pointed X-ray observation of \src\ with a wide spectral coverage
appears worthwhile: If the complex shape is indeed due to a partial
covering absorber, we then need to understand where it could be
located, particularly if \src\ is a low inclination system.

We have observed a Type I X-ray burst; although this is not the first from
this system to be reported \citep{i98}, ours
is the first time-resolved spectral analysis of a burst from \src.  The
spectral cooling we observe is typical of Type I bursts, and can be considered
the definitive evidence that \src\ is a neutron star binary.  The burst
appeared to have peaked at around \sqig 10\% Eddington luminosity,
but with a typical total fluence.
We have approximately 160 s of data after the onset of the burst, and
\src\ clearly had not completed its return to the quiescent state by
the end of this data segment; the duration $\tau$ was about 70 s.
While this duration is relatively long among all X-ray bursts,
it is actually typical of systems with low persistent luminosities
($\gamma$, the ratio of persistent flux to Eddington luminosity,
is about is about 1\% for \src):  $\tau$ ranges from 30s to a few
minutes at $\log \gamma \sim -2$ \citep{v88}.  We conclude that
the \asca\ burst was a typical Type I event.

\citet{d98} have recently suggested a relatively
faint star, star 49, as a possible optical counterpart.  This faintness
may be intrinsic, or geometric: since most of the optical light from
an LMXB is due to reprocessing in the \sqig flat accretion disk,
a high binary inclination can lead to an apparently faint optical
counterpart.  We can comment on this possibility, as the GIS-3 light
curve probably is the most suitable X-ray data for orbital period search
ever obtained for \src.  Since we do not detect orbital modulations,
such as eclipses, dips, or quasi-sinusoidal modulations, we conclude
that \src\ is unlikely to be a high inclination system.

\src\ was seen at X-ray luminosity levels (\sqig 10$^{36}$ ergs\,s$^{-1}$)
typical of X-ray bursters during the \rosat\ and \asca\ observations.
This lends additional support against \src\ being a high-inclination,
Accretion Disc Corona source.  Moreover, we (as well as \citet{i98})
observed what appears to be a typical Type I X-ray burst,
suggesting that we do directly observe the neutron star in \src.  These
provide additional arguments against \src\ being a high inclination system.

If this LMXB is at a low inclination, then a natural explanation for the
optical faintness would be that it is ultra-compact, perhaps similar to
X1820$-$303 in NGC 6624 \citep{a97}.  Since optical
luminosity is dominated by reprocessing, smaller systems tend to be optically
fainter.  We consider this to be a circumstantial evidence for \src\
being an ultracompact binary, joining those in NGC~6624, NGC~6712, and
perhaps NGC~1851 \citep{s87,h96,d96}.

\end{document}